\documentclass[conference]{IEEEtran}
\IEEEoverridecommandlockouts

\usepackage{algorithmic}
\usepackage{textcomp}
\usepackage{amsmath, amssymb, amsthm}
\usepackage{graphicx}
\usepackage{cite}
\usepackage{booktabs}
\usepackage{array}
\usepackage{url}
\usepackage{xcolor}
\usepackage{bm}
\usepackage{booktabs}
\usepackage{multirow}
\usepackage{graphicx}
\usepackage{makecell}
\usepackage{tabularx}

\def\BibTeX{{\rm B\kern-.05em{\sc i\kern-.025em b}\kern-.08em
		T\kern-.1667em\lower.7ex\hbox{E}\kern-.125emX}}
\begin{document}
	
	\title{ISAC-Enabled Grant-Free Uplink via Artificial-Path Delay Modulation}

	\author{Ruiqi Kong~and He Chen\\
		
		\IEEEauthorblockA{Department of Information Engineering, The Chinese University of Hong Kong, Hong Kong, China}
		Email: \{rqkong, he.chen\}@ie.cuhk.edu.hk
		
	}
	
	\maketitle
	
	\begin{abstract}
		This paper proposes an integrated sensing and communication (ISAC)-enabled grant-free uplink framework based on artificial-path delay modulation. A grant-free user equipment (g-UE) conveys uplink information by modulating the delay of a controllable artificial path derived from the scheduled downlink waveform. In contrast to conventional superposition-based schemes with successive interference cancellation, the proposed method enables uplink-downlink coexistence in the delay-sensing domain. By introducing a single weak artificial path confined within the cyclic prefix (CP), the g-UE allows the access point (AP) to decode uplink symbols from CSI perturbations while causing only limited degradation to the scheduled user equipment (s-UE) in the downlink. To support reliable finite-alphabet delay detection under unknown path gain and off-grid leakage, we develop a baseline delay calibration procedure and a normalized matched-filter detector. Results show that reflection power determines the reliability trade-off between the g-UE and the s-UE, whereas the delay step mainly controls the g-UE reliability-efficiency trade-off with little additional impact on the downlink s-UE. Even with an artificial path 15 dB weaker than the scheduled downlink signal, the g-UE achieves lower BER than the s-UE at an effective modulation order of 16-QAM. The proposed framework thus offers a low-complexity, SIC-free, and downlink-friendly solution for grant-free uplink in ISAC systems.
	\end{abstract}
	
	\begin{IEEEkeywords}
		Integrated sensing and communication, grant-free access, delay-domain modulation, artificial path.
	\end{IEEEkeywords}
	
	\section{Introduction}
	Today's wireless systems predominantly rely on orthogonal multiple access (OMA), such as TDMA and OFDMA, where users are separated in time, frequency, or code to avoid mutual interference. This design simplifies resource allocation and receiver implementation, but it may underutilize spectrum when traffic is sporadic, channel conditions are heterogeneous, or a massive number of devices contend for access. To relax strict orthogonality, non-orthogonal multiple access (NOMA) allows multiple users to share the same time--frequency resource via non-orthogonal signaling, most commonly through power-domain superposition and successive interference cancellation (SIC) \cite{saito2013system,ding2017survey,dai2018survey}. By exploiting channel disparities and concurrent transmissions, NOMA can enlarge the achievable rate region relative to OMA and is especially attractive for low-latency and massive-access scenarios \cite{zhao2019nonorthogonal,liu2022optimization,vaezi2019noma}.
	
	However, the practical impact of NOMA has remained limited despite intensive research and 3GPP study \cite{ahmed2025unveiling,yuan2020noma3gpp,yuan2021noma}. A central reason is that its gains are typically coupled with multiuser detection, interference cancellation, and tighter receiver coordination, which raise implementation complexity, power consumption, and deployment cost \cite{vaezi2019noma,yuan2020noma3gpp,yuan2021noma}. In particular, 3GPP investigated NOMA for 5G, yet no NOMA-based solution was ultimately specified for 5G massive machine-type communications (mMTC) support \cite{yuan2020noma3gpp,yuan2021noma}. Meanwhile, much of the ongoing 6G physical-layer discussion still evolves around OFDM-compatible designs rather than a radical departure from the legacy air interface \cite{solaija2024ofdm}. This raises a fundamental question: \textit{can one realize NOMA-like spectrum reuse and grant-free uplink access while preserving the baseline OFDM downlink and avoiding multiuser detection at scheduled users}?
	
	Integrated sensing and communication (ISAC) offers an alternative viewpoint. Beyond using radio resources for data delivery, ISAC endows the access point (AP) with the ability to sense propagation changes from the communication waveform itself \cite{wei2023isacsignals}. In parallel, WiFi sensing standardization, especially IEEE 802.11bf, is pushing CSI-aware sensing toward mainstream wireless systems \cite{ropitault2024wlan}, while commodity CSI tools and subsequent WiFi sensing studies have already demonstrated that fine-grained channel variations can be measured and exploited in practice \cite{halperin2011tool,xie2015precise,tan2022commodity}. Similar sensing-oriented architectural discussions are also emerging in cellular/Open RAN settings \cite{lindenschmitt2025analysis}. These developments suggest that the AP sensing chain can be repurposed not only to observe incidental environmental dynamics, but also to decode intentionally induced and communication-bearing propagation perturbations.
	
	Motivated by this insight, this paper proposes an ISAC-enabled grant-free uplink framework based on artificial-path delay modulation. Rather than transmitting an independent co-channel waveform on top of the scheduled downlink, the grant-free user equipment (g-UE) reuses the scheduled downlink signal and conveys its uplink symbol by controlling the delay of a weak artificial propagation path. This design shifts uplink--downlink coexistence from conventional power-domain superposition to delay-domain sensing. By constraining the artificial path within the cyclic prefix (CP) and keeping its reflection weak, the scheduled user equipment (s-UE) can continue standard OFDM reception without SIC, waveform modification, or hardware changes.
	
	The main contributions of this work are summarized as follows. First, we introduce a downlink-friendly grant-free uplink mechanism in which the g-UE embeds information into the delay position of a controllable artificial path, enabling NOMA-like spectrum reuse through AP-side sensing rather than scheduled-user-side interference cancellation. Second, we develop a CP-constrained delay-position modulation design, where the feasible delay alphabet is jointly determined by the OFDM parameters and the calibrated baseline propagation delay. Third, we formulate symbol detection as a normalized matched-filter search over a fine grid and combine it with baseline delay calibration and reference subtraction to improve robustness under unknown path gain and off-grid leakage. Fourth, we characterize the BER trade-off between the g-UE and the s-UE, showing that the reflection power governs the reliability trade-off between the two links, whereas the delay step mainly controls the g-UE reliability--efficiency trade-off. 
	
	\section{System Overview and Practical Challenges}
	
	\subsection{System Overview}
\begin{figure}
		\centering
		\includegraphics[width=\linewidth]{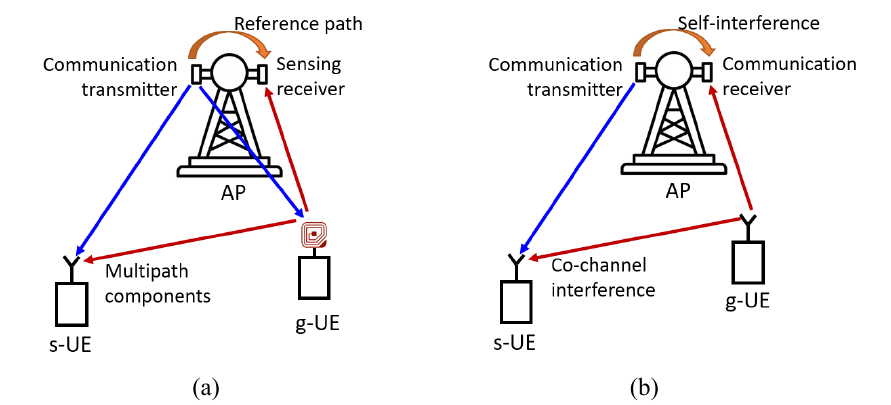}
		\vspace{-1em}
		\caption{(a) System model of ISAC-enabled NOMA. (b) Conventional NOMA}
		\label{fig:system_model}
		\vspace{-1.5em}
	\end{figure}

	As illustrated in Fig.~\ref{fig:system_model}(a), we consider a three-node network consisting of an integrated sensing and communication (ISAC)-capable access point (AP), a scheduled user equipment (s-UE), and a grant-free user equipment (g-UE). 
	The AP operates in a full-duplex manner, simultaneously transmitting a downlink signal to the s-UE while its sensing receiver monitors the channel for grant-free uplink access. The g-UE transmits low-latency uplink data to the AP by dynamically reflecting the ongoing downlink waveform, thereby synthesizing a set of controllable artificial propagation paths that encode its information. The AP's sensing receiver captures the resulting channel variations induced by these artificial paths and decodes the embedded uplink data accordingly.
	
	From the AP's perspective, these reflected components are captured by the sensing receiver and manifested as variations in the composite CSI. By estimating such artificial path variations, the AP can decode the information embedded by the g-UE. From the s-UE's perspective, the reflected signals appear merely as additional multipath components superimposed on the scheduled downlink signal. By properly constraining the reflection power at the g-UE, each reflection contributes only a distinct first-order path, while higher-order reflections are negligible. As such, the impact of the reflected signals on scheduled downlink reception remains minimal.

	We assume the adoption of an OFDM waveform, consistent with standard practical wireless communication systems. The received signal at the AP's sensing receiver is expressed as
	\begin{equation}
		y_{\mathrm{ap}}(t)=h(t)\otimes x(t)+g(t)\otimes x(t)+n(t),
		\label{eq:received_signal}
	\end{equation}
	where $x(t)$ denotes the time-domain OFDM symbol transmitted by the AP at time $t$; $h(t)$ represents the continuous-time channel impulse response~(CIR) of the natural multipath channel between the AP's transmit and receive antennas in the absence of
	reflections from the g-UE; and $n(t)$ is additive noise. The term $g(t) \otimes x(t)$ captures the contribution of the g-UE's reflection, where the continuous-time CIR $g(t)$ comprises the controllable artificial multipath components that encode the g-UE's data.

	Ideally, the natural CIR and the artificial CIR can be expressed as
	\begin{equation}
		h(t)=\sum\nolimits_{\ell=1}^{L_1}\alpha_{\ell}(t)e^{-j2\pi f_c\tau_{\ell}(t)}\delta\bigl(t-\tau_{\ell}(t)\bigr),
	\end{equation}
	and
	\begin{equation}
		g(t)=\sum\nolimits_{k=1}^{L_2}\beta_k(t)e^{-j2\pi f_c\tau_k(t)}\delta\bigl(t-\tau_k(t)\bigr),
	\end{equation}
	where $L_1$ and $L_2$ denote the numbers of natural and artificial paths, respectively, $\alpha_{\ell}(t)$ and $\beta_k(t)$ are the corresponding complex path gains, $\tau_{\ell}(t)$ and $\tau_k(t)$ denote the path delays, $f_c$ is the carrier frequency, and $\delta(\cdot)$ is the Dirac delta function. The g-UE's information can be embedded into the artificial-path parameters, such as the complex gains $\beta_k(t)$ and/or the path delays $\tau_k(t)$.
	
	\begin{figure}[]
		\centering
		\includegraphics[width=0.95\linewidth]{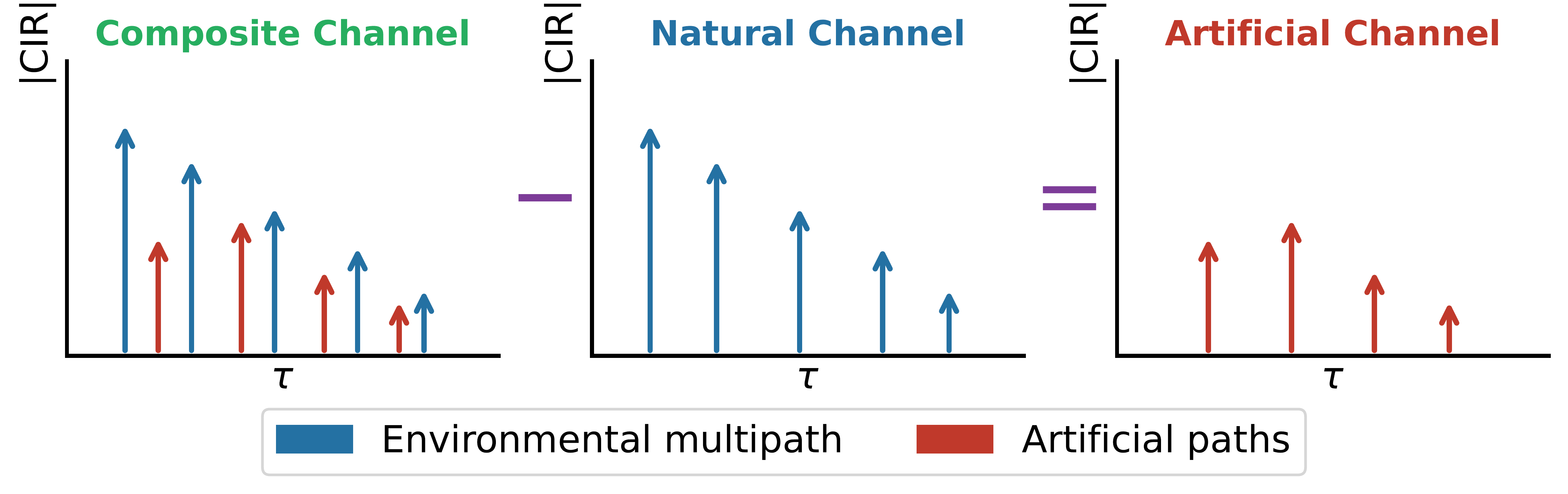}
		\vspace{-1em}
		\caption{Artificial-path extraction at the AP sensing receiver. The AP estimates the composite channel, removes the reference natural channel, and decodes the information embedded in the artificial channel component.}
		\label{fig:artificial_path_extraction}
		\vspace{-1.5em}
	\end{figure}
	
	Under this model, the AP estimates the composite channel using pilot symbols embedded in the OFDM waveform. The artificial-path extraction can be carried out equivalently in either the frequency domain or the delay domain. In particular, the AP may directly subtract a reference estimate of the natural channel from the composite channel estimate in the frequency domain to isolate the artificial component. Alternatively, it may transform the estimated channel into the delay domain and perform the same subtraction on the corresponding CIR representation. These two implementations are equivalent up to a Fourier transform. After removing the natural channel contribution, the AP obtains the artificial channel component associated with the g-UE and recovers the g-UE's grant-free information. This artificial-path extraction procedure is illustrated in Fig.~\ref{fig:artificial_path_extraction}.

	Compared with conventional power-domain NOMA, the proposed sensing-assisted non-orthogonal access provides a different interference-management mechanism. In conventional NOMA, multiple information-bearing signals are directly superimposed in the same time-frequency resources, and the receiver typically relies on multiuser detection techniques such as successive interference cancellation (SIC). In contrast, the proposed framework embeds the grant-free information into controllable channel variations. At the s-UE, the g-UE-induced reflections behave as additional multipath components rather than as an independently generated co-channel waveform. As long as the artificial paths remain within the cyclic prefix (CP) duration, OFDM subcarrier orthogonality is preserved, and the scheduled downlink reception is only marginally affected.
	
	\begin{figure}[]
		\centering
		\includegraphics[width=0.9\linewidth]{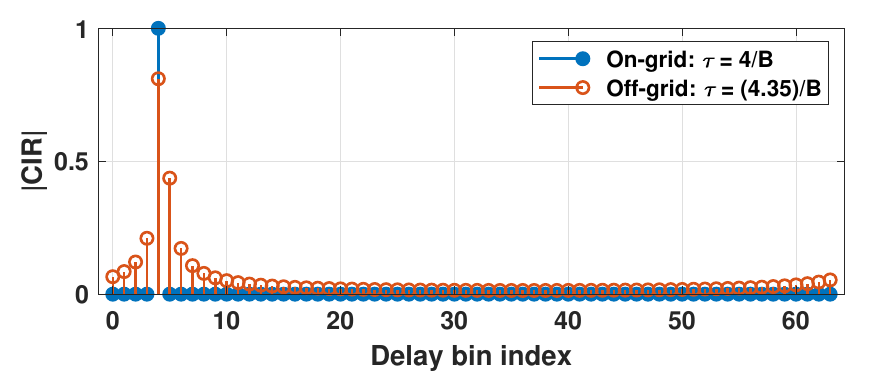}
		\vspace{-1em}
		\caption{Illustration of delay-domain leakage. An on-grid artificial path concentrates its energy in a single delay bin, whereas an off-grid path spreads energy across multiple neighboring bins.}
		\label{chanel_leakage}
		\vspace{-1.5em}
	\end{figure}
	
	\subsection{Practical Challenges}
	\label{sec:challenges}
	
	Although the above mechanism provides a practical path toward grant-free low-latency access, several implementation challenges must be addressed.
	
	\subsubsection{Off-Grid Delay Effect}
	
	Physical path delays are continuous-valued and generally do not coincide with the discrete delay grid imposed by the finite observation bandwidth $B = N\Delta f$. As illustrated in
	Fig.~\ref{chanel_leakage}, an on-grid path (e.g., $\tau = 4/B$) produces a clean, concentrated response, whereas an off-grid path (e.g., $\tau = 4.35/B$) spreads its energy across multiple neighboring bins rather than concentrating on a single impulse. 
	This leakage makes it difficult to estimate artificial components and introduces mismatches when assuming idealized discrete delay models. Accurately recovering the artificial-path parameters, such as the complex gains $\beta_k$ and delays $\tau_k$, in the presence of this leakage is therefore the first core challenge.

	\subsubsection{Minimal Impact on s-UE Transmission}
	The second challenge concerns the impact of the artificial paths on the s-UE's downlink reception. Each reflection introduced by the g-UE adds an extra propagation path to the channel seen by the s-UE, effectively augmenting the natural multipath profile.
	While a single weak artificial path causes only a mild perturbation, introducing multiple paths or paths with non-negligible power progressively increases the frequency
	selectivity of the composite channel. This manifests as deeper frequency-domain fading across OFDM subcarriers, which degrades channel quality at the s-UE and ultimately impairs data detection performance. The g-UE must therefore carefully control both the number of artificial paths and their individual power levels to keep the induced frequency-selective distortion within an acceptable range, while still conveying sufficient information
	to the AP. Striking this balance is the second core challenge of the proposed system.

	\section{Artificial-Path Delay Modulation and Demodulation}
	
	To address the two practical challenges identified in Section~\ref{sec:challenges}, we propose a delay-only artificial-path modulation scheme. The key idea is to let the g-UE encode each information symbol by selecting the delay of a single controllable artificial path, while keeping the reflected signal weak and structurally simple. 
	As an initial step, we consider a simple instantiation that employs only a single path and a single modulation dimension. More advanced extensions, such as multi-path designs or joint amplitude-phase-delay modulation, are certainly possible, but we leave them to future work. This simplified setting allows us to isolate and reveal the fundamental trade-offs of the proposed framework.

	From the AP's perspective, this design also simplifies detection. After the natural channel contribution is removed, the AP only needs to identify which delay hypothesis best matches the residual artificial channel. Therefore, the detection problem becomes a one-dimensional delay-domain classification problem. This structure directly addresses the off-grid leakage challenge: instead of relying on a coarse DFT-bin search, the AP evaluates a matched-filter metric using continuous-delay steering vectors, which explicitly capture the phase pattern of off-grid paths.
	
	\subsection{Design Rationale}
	The proposed modulation design follows two principles. First, the delay of the artificial path must remain within the CP-safe region of the OFDM system. Therefore, the available controllable delay range is fundamentally bounded, and any practical modulation design must explicitly account for this physical constraint. Second, the absolute delay of the reflected path is not fully controllable by the g-UE. The observed delay includes an uncontrollable propagation component determined by the AP--g-UE--AP path geometry. The modulation should therefore be defined relative to a calibrated baseline delay, so that each symbol corresponds to a quantized delay offset from the zero-controllable-delay state. 
	
	\subsection{Baseline Calibration}
	\label{sec:calibration}
	
	Before data transmission, the AP estimates the baseline delay of the artificial path through a two-step calibration procedure. In the first reference packet, the g-UE remains inactive, so the AP observes only the natural environment channel:
	\begin{equation}
		\mathbf{H}^{(0)} = \mathbf{h}_{\mathrm{env}} + \mathbf{w}^{(0)},
	\end{equation}
	where $\mathbf{h}_{\mathrm{env}}$ denotes the environment channel component and $\mathbf{w}^{(0)}$ denotes noise. In the second reference packet, the g-UE is activated with zero controllable delay offset, yielding
	\begin{equation}
		\mathbf{H}^{(1)} = \mathbf{h}_{\mathrm{env}} + \beta \mathbf{a}(\tau_{\mathrm{base}}) + \mathbf{w}^{(1)},
	\end{equation}
	where $\beta$ is the complex gain of the artificial path, $\tau_{\mathrm{base}}$ is the baseline delay of the path, and $\mathbf{a}(\tau)$ denotes the delay steering vector whose $n$-th entry is $e^{-j2\pi n\Delta f\tau}$.
	
	By subtracting the two observations, the environment component is removed:
	\begin{equation}
		\mathbf{y}_{\mathrm{cal}}
		=
		\mathbf{H}^{(1)}
		-
		\mathbf{H}^{(0)}
		=
		\beta \mathbf{a}(\tau_{\mathrm{base}})
		+
		\mathbf{v}_{\mathrm{cal}},
		\label{eq:calibration_observation}
	\end{equation}
	where $\mathbf{v}_{\mathrm{cal}} = \mathbf{w}^{(1)} -
	\mathbf{w}^{(0)}$ is the combined noise term.

	The baseline delay is estimated using a normalized matched-filter criterion, which can be derived from a maximum-likelihood formulation with an unknown complex path gain. 
	Under additive white Gaussian noise, jointly estimating the delay and gain is equivalent to minimizing the residual energy:
	$
	(\hat{\tau},\hat{\beta})
	=
	\arg\min_{\tau,\beta}
	\left\|
	\mathbf{y}_{\mathrm{cal}}
	-
	\beta\mathbf{a}(\tau)
	\right\|^2.
	$
	For a fixed candidate delay $\tau$, the least-squares estimate of the gain is
	\begin{equation}
		\hat{\beta}(\tau)
		=
		\frac{
			\mathbf{a}^{H}(\tau)\mathbf{y}_{\mathrm{cal}}
		}{
			\|\mathbf{a}(\tau)\|^2
		}.
	\end{equation}
	Substituting $\hat{\beta}(\tau)$ back into the residual minimization problem shows that the optimal delay is obtained by maximizing the normalized projection energy. Therefore, the AP estimates the baseline delay as
	\begin{equation}
		\hat{\tau}_{\mathrm{base}}
		=
		\arg\max_{\tau}
		\frac{
			\left|
			\mathbf{a}^{H}(\tau)
			\mathbf{y}_{\mathrm{cal}}
			\right|^2
		}{
			\|\mathbf{a}(\tau)\|^2
		}.
	\end{equation}
	The search is performed over a fine continuous delay grid rather than over coarse DFT delay bins. By evaluating continuous-delay steering vectors, the matched-filter estimator directly captures the phase response of off-grid paths and mitigates leakage-induced delay bias.
	
	This calibration establishes the reference point for all subsequent modulation and detection operations.
	
	\subsection{CP-Constrained Delay Alphabet Construction}
	\label{sec:alphabet}
	
	After calibration, the AP constructs a CP-safe delay alphabet. Let $\Delta_\tau$ denote the delay quantization unit, and let $M$ denote the modulation order. The transmitted symbol in packet $i$ is represented by an integer
	$
	k_i \in \{0,1,\dots,M-1\}.
	$
	The corresponding artificial path delay is set to
	$
	\tau_i = \hat{\tau}_{\mathrm{base}} + k_i \Delta_\tau.$

	Since the total path delay must remain within the CP-safe region, the maximum induced delay must satisfy
	\begin{equation}
		\hat{\tau}_{\mathrm{base}} + (M-1)\Delta_\tau
		\leq T_{\mathrm{CP}} - T_{\mathrm{guard}},
	\end{equation}
	where $T_{\mathrm{CP}}$ is the cyclic prefix duration and
	$T_{\mathrm{guard}}$ is a design margin that accounts for
	estimation uncertainty and implementation nonidealities.
	This constraint directly bounds the admissible modulation
	order:
	\begin{equation}
		M \leq 1 + \left\lfloor
		\frac{T_{\mathrm{CP}} - T_{\mathrm{guard}}
			- \hat{\tau}_{\mathrm{base}}}{\Delta_\tau}
		\right\rfloor.
	\end{equation}
	This expression reveals a fundamental trade-off: a finer
	quantization step $\Delta_\tau$ supports a higher modulation
	order but demands more accurate delay estimation, while a
	larger baseline delay $\hat{\tau}_{\mathrm{base}}$ reduces
	the available delay range and thus limits spectral efficiency.
	The delay alphabet is therefore not chosen heuristically,
	but is determined entirely by the OFDM system parameters
	and the estimated propagation environment.
	
	\subsection{Delay-Domain Symbol Detection}
	\label{sec:detection}
	
	Each data packet $i$ yields a composite channel observation
	at the AP:
	\begin{equation}
		\mathbf{H}_i = \mathbf{h}_{\mathrm{env},i}
		+ \beta_i\,\mathbf{a}(\tau_i) + \mathbf{w}_i,
	\end{equation}
	where $\mathbf{h}_{\mathrm{env},i}$ is the environment channel,
	$\beta_i$ is the artificial-path complex gain, and
	$\mathbf{w}_i$ is additive noise. The AP removes the
	environment contribution using a reference estimate channel
	$\hat{\mathbf{h}}_{\mathrm{env},i}$ maintained via periodic
	calibration or background tracking, yielding
	\begin{equation}
		\mathbf{y}_i = \mathbf{H}_i - \hat{\mathbf{h}}_{\mathrm{env},i}
		\approx \beta_i\,\mathbf{a}(\tau_i) + \mathbf{v}_i,
		\label{eq:data_residual_observation}
	\end{equation}
	where $\mathbf{v}_i$ subsumes residual environment mismatch
	and noise. Symbol detection is then performed on $\mathbf{y}_i$.
	
	The detection rule follows the same unknown-gain maximum-likelihood principle used in the calibration stage. Given the calibrated delay alphabet
	$\mathcal{T} = \{\hat{\tau}_{\mathrm{base}}
	+ k\Delta_\tau : k = 0,\dots,M-1\}$,
	the AP evaluates the normalized matched-filter metric for
	each symbol hypothesis:
	\begin{equation}
		\Lambda_i(k)
		= \frac{\left|\mathbf{a}_k^{H}\,\mathbf{y}_i\right|^2}
		{\|\mathbf{a}_k\|^2},
		\qquad k = 0,\dots,M-1,
	\end{equation}
	where $\mathbf{a}_k \triangleq
	\mathbf{a}(\hat{\tau}_{\mathrm{base}} + k\Delta_\tau)$. Therefore, the AP detects the transmitted symbol by selecting the delay hypothesis with the largest matched-filter output:
	\begin{equation}
		\hat{k}_i = \arg\max_{k}\;\Lambda_i(k),
		\label{eq:delay_alphabet_detection}
	\end{equation}
	The corresponding bit block is then recovered through inverse symbol mapping.
	
	This detector directly addresses the off-grid leakage challenge. The candidate delays in Eq.~\eqref{eq:delay_alphabet_detection} are not restricted to coarse DFT delay bins; instead, they are constructed from the calibrated continuous baseline delay $\hat{\tau}_{\mathrm{base}}$ and the designed delay spacing $\Delta_\tau$. 
	Meanwhile, the s-UE observes at most one weak CP-contained additional multipath component, rather than multiple artificial echoes or an independently superimposed co-channel waveform. Thus, the proposed detection rule preserves simple grant-free decoding at the AP while maintaining limited perturbation to the scheduled downlink reception.

	\section{Numerical Evaluation}
	\label{sec:Evaluations}
	
	\subsection{Setup and Definitions}
	
	We conduct three simulations to evaluate the proposed framework from two complementary perspectives: (i)~the precision of AP-side delay estimation, and (ii)~the BER trade-offs between the g-UE's uplink and the scheduled downlink under delay-position modulation. The key parameters shared across Evaluations are summarized in Table~\ref{tab:key_exp_params}.

	\begin{table}[t]
		\centering
		\caption{Key Evaluational parameters.}
		\label{tab:key_exp_params}
		\scriptsize
		\setlength{\tabcolsep}{3pt}
		\renewcommand{\arraystretch}{1.08}
		\resizebox{\columnwidth}{!}{
			\begin{tabular}{l l l c}
				\toprule
				\textbf{Parameter} & \textbf{Symbol} & \textbf{Value} & \textbf{Eval.} \\
				\midrule
				CSI subcarriers 
				& \(N\) 
				& 64 
				& 1--3 \\
				System bandwidth 
				& \(B\) 
				& \(20\,\mathrm{MHz}\) 
				& 1--3 \\
				CP length 
				& \(N_{\mathrm{CP}}\) 
				& 16 samples 
				& 2--3 \\
				Base delay of g-UE
				& \(\tau_{\mathrm{base}}\) 
				& 3 samples 
				& 2--3 \\
				Delay search range 
				& \([\tau_{\min}, \tau_{\max}]\) 
				& \(50\)--\(800\,\mathrm{ns}\) 
				& 1 \\
				Delay grid step 
				& \(\Delta\tau_{\mathrm{grid}}\) 
				& \(1\,\mathrm{ns}\) 
				& 1 \\
				s-UE modulation 
				& -- 
				& 16-QAM 
				& 2--3 \\
				Monte Carlo trials 
				& \(N_{\mathrm{trials}}\) 
				& \(10^5\) 
				& 1--3 \\
				Alphabet level rule 
				& \(L\) 
				& \(2^{\lfloor \log_2(10/\mathrm{step}) \rfloor}\) 
				& 2--3 \\
				\bottomrule
			\end{tabular}
		}
		\vspace{-1em}
	\end{table}

	It is important to note that the SNR definitions differ between Evaluation~1 and Evaluations~2--3, reflecting the distinct evaluation objectives of each group.
	In Evaluation~1, SNR is defined as the ratio of artificial-path power to noise power, i.e., the observation model is
	$
	\mathbf{y} = \mathbf{a}(\tau)\beta + \mathbf{n},
	$
	which directly characterizes the quality of the AP-side delay estimation.
	In Evaluations~2--3, SNR is defined as the ratio between the environmental signal power and the noise power. The artificial-path strength is controlled separately by the reflection power ratio \ensuremath{\rho}, measured in dB relative to the environmental signal power. This separation allows us to independently examine the effect of environmental SNR and artificial reflection strength.
	
	\subsection{Evaluation 1: Delay Estimation Accuracy vs.\ SNR}
	
	\begin{figure}[t]
		\centering
		\includegraphics[width=0.8\linewidth]{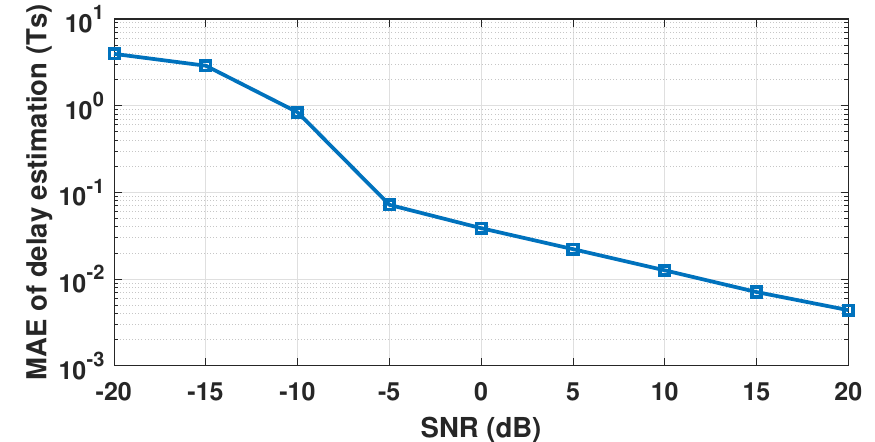}
		\vspace{-1em}
		\caption{Mean absolute delay estimation error versus SNR under the
			AP-side single-path observation model.}
		\label{fig:exp1_tau_error}
		\vspace{-1em}
	\end{figure}

	\subsubsection{Methodology}
	This Evaluation evaluates the delay estimation accuracy of the normalized matched-filter estimator. For each SNR point, a single-path observation $\mathbf{y} = \mathbf{a}(\tau)\beta + \mathbf{n}$ is generated with a randomly drawn delay
	$\tau \in [\tau_{\min}, \tau_{\max}]$ for each trial. The AP performs a one-dimensional grid search over the delay axis and selects the delay hypothesis that maximizes the normalized matched-filter response. The mean absolute error~(MAE) is
	computed over $N_{\mathrm{trials}} = 10^5$ independent Monte Carlo runs per SNR point.
	
	\subsubsection{Results and Analysis}

	Fig.~\ref{fig:exp1_tau_error} shows that the delay estimation MAE decreases monotonically with SNR. At low SNR, noise-dominated matched-filter outputs lead to large errors on the order of several sampling periods. As SNR increases, the artificial path becomes more distinguishable, and the matched-filter response concentrates around the true delay, causing the MAE to drop rapidly. 
	The transition from coarse to accurate delay estimation occurs mainly from the low-to-medium SNR regime. The MAE falls below \ensuremath{10^{-1}T_s} around \ensuremath{-5\,\mathrm{dB}} and reaches the \ensuremath{10^{-2}T_s} level in the medium-to-high SNR range. This behavior is consistent with the threshold effect of maximum-likelihood-type estimators and confirms that the proposed estimator provides sufficiently fine delay resolution for reliable delay-alphabet design.
	
	\subsection{Evaluation 2: BER Trade-off under Reflection Power}

	\begin{table}[t]
		\centering
		\caption{BER performance of the g-UE and s-UE under different reflection powers with $step=0.15$ and $L=64$. The baseline row reports the s-UE BER without artificial-path modulation.}
		\label{tab:exp2_fixed_step_data}
		\small
		\setlength{\tabcolsep}{3pt}
		\renewcommand{\arraystretch}{1.08}
		\resizebox{\columnwidth}{!}{
			\begin{tabular}{c|ccc|ccc}
				\toprule
				\multirow{2}{*}{\ensuremath{\rho} (dB)}
				& \multicolumn{3}{c|}{g-UE BER}
				& \multicolumn{3}{c}{s-UE BER} \\
				\cmidrule(lr){2-4} \cmidrule(lr){5-7}
				& 0 dB & 10 dB & 20 dB 
				& 0 dB & 10 dB & 20 dB \\
				\midrule
				-20 & 0.4690 & 0.2255 & 0.0097  & 0.3621 & 0.1112 & 0.0007  \\
				-15 & 0.4054 & 0.0343 & 0.0015  & 0.3622 & 0.1127 & 0.0010 \\
				-10 & 0.2313 & 0.0087 & 0  & 0.3632 & 0.1179 & 0.0025  \\
				-5 & 0.0353 & 0.0017 & 0  & 0.3633 & 0.1331 & 0.0108  \\
				\midrule
				Baseline & -- & -- & --  & 0.3622 & 0.1107 & 0.0006  \\
				\bottomrule
			\end{tabular}
		}
		\vspace{-1em}
	\end{table}

	\subsubsection{Methodology}
	This Evaluation evaluates how the reflection power affects both the g-UE uplink reliability and the scheduled s-UE downlink reception. The delay quantization step is fixed at \ensuremath{step = 0.15}, corresponding to \ensuremath{\Delta_\tau = step/B}, which yields a modulation alphabet of \ensuremath{L = 64} levels. The reflection power \ensuremath{\rho} is swept over \ensuremath{\{-20, -15, -10, -5\}\,\mathrm{dB}} relative to the environment signal power. For each configuration, the g-UE BER and s-UE BER are jointly evaluated at SNR values of 0, 10, 20, and 30 dB. The baseline row in Table~\ref{tab:exp2_fixed_step_data} reports the s-UE BER without artificial-path modulation, serving as a reference for the undisturbed downlink.
	
	\subsubsection{Results and Analysis}

	Table~\ref{tab:exp2_fixed_step_data} demonstrates a clear trade-off between g-UE detectability and s-UE transparency. As the reflection power increases from \ensuremath{-20\,\mathrm{dB}} to \ensuremath{-5\,\mathrm{dB}}, the g-UE BER decreases substantially. For example, at 10 dB SNR, the g-UE BER drops from 0.2255 to 0.0017, and at 20 dB SNR, it decreases from 0.0097 to zero observed errors. This improvement occurs because a stronger artificial path increases the effective SNR at the AP's matched-filter output, making the delay hypotheses easier to distinguish.
	
	The s-UE, however, experiences a mild degradation as the reflection power increases. At 20 dB SNR, the s-UE BER rises from 0.0007 at \ensuremath{\rho=-20\,\mathrm{dB}} to 0.0108 at \ensuremath{\rho=-5\,\mathrm{dB}}. This degradation is caused by the stronger artificial reflection, which introduces more pronounced frequency-selective distortion into the scheduled downlink signal. The effect is especially visible in the moderate-to-high SNR regime, where the residual distortion rather than thermal noise becomes the dominant impairment.
	
	The baseline row provides a reference for the s-UE BER without artificial-path modulation. When the reflection power is sufficiently low, e.g., \ensuremath{\rho=-20\,\mathrm{dB}} or \ensuremath{\rho=-15\,\mathrm{dB}}, the s-UE BER remains close to the baseline across all SNR values. This confirms that weak artificial paths can remain nearly transparent to the scheduled downlink while still enabling reliable g-UE detection at moderate and high SNR. Entries reported as zero indicate that no bit errors were observed in the corresponding simulations.
	
	\subsection{Evaluation 3: BER Trade-off under Delay Step}

	\begin{table}[t]
		\centering
		\caption{BER performance of the g-UE and s-UE under different delay steps and constellation sizes with reflection power fixed at -15 dB.}
		\label{tab:exp3_fixed_refl_data}
		\small
		\setlength{\tabcolsep}{3pt}
		\renewcommand{\arraystretch}{1.08}
		\resizebox{\columnwidth}{!}{
			\begin{tabular}{cc|ccc|ccc}
				\toprule
				\multirow{2}{*}{Step} & \multirow{2}{*}{L}
				& \multicolumn{3}{c|}{g-UE BER}
				& \multicolumn{3}{c}{s-UE BER} \\
				\cmidrule(lr){3-5} \cmidrule(lr){6-8}
				& & 0 dB & 10 dB & 20 dB 
				& 0 dB & 10 dB & 20 dB \\ 
				\midrule
				0.10 & 64 & 0.3940 & 0.0445 & 0.0057  & 0.3626 & 0.1130 & 0.0010 \\ 
				0.20 & 32 & 0.3921 & 0.0406 & 0.0009  & 0.3627 & 0.1128 & 0.0010 \\ 
				0.30 & 32 & 0.4002 & 0.0314 & 0  & 0.3625 & 0.1128 & 0.0010 \\ 
				0.40 & 16 & 0.3913 & 0.0254 & 0  & 0.3626 & 0.1127 & 0.0010 \\ 
				\bottomrule
			\end{tabular}
		}
		\vspace{-1em}
	\end{table}

	\subsubsection{Methodology}
	
	Reflection power is fixed at \ensuremath{\rho = -15\,\mathrm{dB}}, and the delay quantization step \ensuremath{step} is varied. The modulation order is determined by \ensuremath{L = 2^{\lfloor\log_2(10/step)\rfloor}}, with \ensuremath{\Delta_\tau = step/B}. As shown in Table~\ref{tab:exp3_fixed_refl_data}, each \ensuremath{(step, L)} pair is explicitly listed to facilitate direct comparison between delay-domain symbol efficiency and detection reliability. For each configuration, both the g-UE BER and s-UE BER are evaluated at SNR values of 0, 10, 20, and 30 dB.
	
	\subsubsection{Results and Analysis}
	
	Table~\ref{tab:exp3_fixed_refl_data} shows that the delay step mainly affects the g-UE BER, while the s-UE BER remains nearly unchanged across different delay alphabet configurations. For the g-UE, larger delay steps generally improve the BER performance, especially at moderate and high SNR. At 10 dB SNR, the g-UE BER decreases from 0.0445 for \ensuremath{step=0.10} to 0.0254 for \ensuremath{step=0.40}. 
	This trend is due to the increased separation between adjacent delay-domain steering vectors. A larger \ensuremath{\Delta_\tau} makes neighboring delay hypotheses more distinguishable at the AP's matched-filter detector, thereby reducing the probability of delay-index detection errors. In contrast, the s-UE BER remains almost invariant across the tested delay steps. This indicates that changing the delay-domain alphabet has little impact on the scheduled downlink when the reflection power is fixed.
	
	Notably, for \ensuremath{step\geq 0.2} and \ensuremath{SNR\geq 10} dB, the proposed delay-domain modulation achieves lower g-UE BER than the scheduled 16-QAM downlink while maintaining a modulation order no smaller than that of the s-UE for \ensuremath{L\geq 16}. This is because the AP does not decode the g-UE information from a conventional amplitude/phase constellation. Instead, it performs delay-domain hypothesis testing using the channel response across multiple subcarriers. These results show that the proposed delay-domain access mechanism does not merely obtain reliability by sacrificing modulation order; rather, it can simultaneously provide competitive symbol efficiency and improved BER performance when the delay spacing is properly selected. Entries reported as zero indicate that no bit errors were observed in the corresponding simulations.

	\section{Conclusions and Future Work}
	This paper presented an ISAC-enabled grant-free access framework in which a grant-free user equipment (g-UE) embeds uplink data into the delay of a single CP-contained artificial path created from the scheduled downlink waveform. This allows the AP to decode it via CSI sensing, without requiring SIC or hardware modification at the scheduled user equipment (s-UE). 
	A cyclic prefix (CP)-constrained delay-position modulation design, combined with baseline delay calibration and a normalized matched-filter detector, enables reliable finite-alphabet delay decoding under unknown path gain and off-grid leakage. 
	Even with an artificial path 15 dB weaker than the downlink signal, the g-UE achieves lower BER than the s-UE at 16-QAM, demonstrating a low-complexity, SIC-free, and downlink-friendly solution for grant-free access in ISAC systems. 
	
	Looking ahead, several promising directions remain for future work, including the incorporation of multiple artificial paths, joint modulation of path amplitude and phase together with delay, and the development of more advanced detectors to further enhance spectral efficiency and reliability.
	
	\bibliographystyle{IEEEtran}
	\bibliography{ref.bib}
\end{document}